\documentclass[prc,twocolumn,letterpaper,10pt,twoside,tightenlines,nofootinbib,]{revtex4}

\usepackage{amsmath}
\usepackage{amsfonts}
\usepackage{amssymb}
\usepackage{latexsym}
\usepackage{hyperref}
\usepackage{graphicx,subfigure}
\usepackage{epsfig}
\usepackage[usenames]{color}
\usepackage[letterpaper,left=2.cm,right=2.cm,top=2.5cm,bottom=2.5cm]{geometry}

\begin{document}

\newcommand{\be}{\begin{equation}}
\newcommand{\ee}{\end{equation}}
\newcommand{\beqa}{\begin{eqnarray}}
\newcommand{\eeqa}{\end{eqnarray}}
\newcommand{\beqar}{\begin{eqnarray*}}
\newcommand{\eeqar}{\end{eqnarray*}}

\title{Non-Axisymmetric Instability of Rotating Black Holes in Higher Dimensions}

\author{Gavin S. Hartnett and Jorge E. Santos}
\affiliation{Department of Physics, UCSB, Santa Barbara, CA 93106, USA}
\date{\today}

\begin{abstract} \noindent
We calculate the scalar-gravitational quasi-normal modes of equal angular momenta Myers-Perry black holes in odd dimensions. We find a new bar-mode (non-axisymmetric) classical instability for $D \ge 7$. These black holes were previously found to be unstable to axisymmetric perturbations for spins very near extremality. The bar-mode instability we find sets in at much slower spins, and is therefore the dominant instability of these black holes. This instability has important consequences for the phase diagram of black holes in higher dimensions.
\end{abstract}

\maketitle

\noindent
\emph{Introduction:} 
Black holes in four spacetime dimensions are remarkably featureless. The Kerr metric is the unique stationary and axisymmetric solution to the vacuum Einstein equations, and is completely characterized by just two parameters, its mass and angular momentum \cite{uniqueness}. In contrast, General Relativity in higher dimensions allows for a fantastic diversity of different asymptotically flat black objects. In addition to black holes of spherical topology, there are also black rings \cite{Emparan:2001wn}, black Saturns \cite{Elvang:2007rd}, systems of bicycling black rings \cite{Elvang:2007hs}, and so on \footnote{For a recent and comprehensive review, see \cite{garyhorowitz}}. In fact, it has been shown \cite{Emparan:2007wm} that in marked contrast to the uniqueness of the Kerr metric in four dimensions, in higher dimensions there are many black holes of a given mass and set of angular momenta. Perhaps more exotically, in \cite{Emparan:2009vd}, it has been argued that an \emph{infinite} number of black holes can exist with the same set of conserved asymptotic charges! Understanding the phase space of black hole solutions, as well as their stability, is currently a very active program of research.
\\
\noindent\indent In this Letter we report progress in this direction by studying the classical stability of Myers-Perry (MP) black holes, which are the generalization of the Kerr solution to higher dimensions \cite{Myers:1986un}. In $D$ spacetime dimensions these solutions are characterized by their mass and $\lfloor (D-1)/2 \rfloor$ angular momenta parameters. The properties of these solutions strongly depends on both the dimension of spacetime as well as the angular momenta. For example, for $D \ge 6$,  when none of the angular momenta vanish there is an extremal limit and the black hole cannot be made to rotate arbitrarily fast, just as in four dimensions. However, when at least one of the angular momenta vanishes, there is no extremal limit and the remaining angular momenta can be taken to be arbitrarily large.
\\
\noindent\indent The lack of an extremal limit suggests that these black holes might be unstable. Emparan and Myers \cite{Emparan:2003sy} showed that in the limit where $n$ angular momenta are taken to be arbitrarily large, the horizon `pancakes' out to have topology $\mathbb{R}^{2n} \times S^{D-2n-2}$. Black holes with this horizon shape  are known as black branes, and were famously shown to be unstable by Gregory and Laflamme \cite{Gregory:1993vy}. Therefore, Emparan and Myers conjectured that MP black holes should also be unstable, at least in certain fast-spinning regions of the parameter space. Although the comparison to the Gregory-Laflamme instability was made in the limit of infinite rotation, it was expected that the instability would set for finite, sufficiently rapid rotation.
\\
\noindent\indent Since this conjecture, there has been much work on investigating the stability of MP black holes in various dimensions and for various configurations of the angular momenta parameters. One of the most studied cases has been in odd dimensions with all the angular momenta equal. This is because in this limit the metric becomes cohomogeneity-1, which is to say that it depends non-trivially only on the radial coordinate, and therefore the linearized Einstein equations form a coupled system of ODE's. For generic rotations, the perturbation equations necessarily involve PDE's, greatly complicating the stability analysis. The fact that there exists a cohomogeneity-1 MP metric is rather remarkable, as the only other case in which the linearized Einstein equations are known to separate on a rotating black hole background is Kerr, as shown by Teukolsky \cite{Teukolsky:1973ha}.
\\
\noindent\indent Although the perturbation equations become more tractable for equal angular momenta, in this case the spins cannot be made arbitrarily large, and a priori it is not clear whether the black hole would be able to rotate at a sufficiently rapid speed to become unstable. A precise definition of sufficiently rapid was given in \cite{Dias:2010eu}, who formulated an ultraspinning condition based upon black hole thermodynamics. In odd $D \ge 7$ this ultraspinning condition can be satisfied for the equal angular momenta case, and therefore these black holes might be unstable. Indeed, it was found that very near extremality these black holes were unstable to perturbations that do not break rotational symmetries.\footnote{This instability was first found in $D=9$  \cite{Dias:2010eu}, but upon further investigation (motivated by the study of the relationship of near-horizon instabilities to instabilities of the full geometry in Ref. \cite{Durkee:2010ea}) the instability was found to also exist in $D=7$.} It is expected that this instability persists for all odd $D \ge 7$. 
\\
\noindent\indent No instability was found in $D=5$, and in fact this was not unexpected since the ultraspinning condition is only possible in $D\ge7$. Further evidence for stability in $D=5$ for the case of equal angular momenta came from the linear perturbative analysis of Ref. \cite{Murata:2008yx}. The study of axisymmetric perturbations has been extended to configurations with a single non-vanishing spin \cite{Dias:2009iu}, and more general configurations \cite{Dias:2011jg}. Instabilities have been found in all $D \ge 6$. These instabilities are very important for the phase diagram of black holes in higher dimensions, because at the threshold of instability, the perturbations are time-independent and therefore correspond to a new family of stationary, axisymmetric black holes branching off of the MP family. 
\\
\noindent\indent Thus far the discussion has been restricted to axisymmetric perturbations. Of course a full analysis of the stability of any physical system should include all possible perturbations; therefore we now turn to discuss non-axisymmetric perturbations. In an impressive series of numerical simulations, it was found that singly-spinning MP black holes were unstable to non-axisymmetric instabilities, first in $D=5$ \cite{Shibata:2009ad}, and then in $D=6,7,8$ \cite{Shibata:2010wz}. These results strongly suggest that black holes might be much more sensitive to non-axisymmetric perturbations than to axisymmetric perturbations, since $D=5$ black holes are stable to axisymmetric perturbations, but unstable to non-axisymmetric ones. It might therefore be expected that in higher $D$ non-axisymmetric instabilities set in for smaller spins than the axisymmetric ones.
\\
\noindent\indent The above results motivated the work presented here. We study non-axisymmetric perturbations of higher dimensional black holes. We restrict our attention to the equal angular momenta case (and therefore odd $D$), due to the simplification that occurs when the metric is cohomogeneity-1. We find that in odd $D \ge 7$, equal angular momenta MP black holes are unstable to non-axisymmetric perturbations, and that these instabilities set in for much smaller rotations than the previously discovered axisymmetric ultraspinning instabilities \cite{Dias:2010eu}. Our analysis explores the full spectra of scalar-gravitational perturbations, in particular its quasi-normal mode frequencies. We studied the cases of $D = 5,7,9,11,13,15$, and found instabilities for $D \ge 7$. We expect these instabilities to persist for  all higher odd $D$. 
\\ \\
\noindent \emph{Methodology and Results:} The equal angular momenta Myers-Perry black hole line element is
\beqa
\bar{ds}^2 &=& -f(r)^2dt^2 +g(r)^2dr^2  \\
&+& h(r)^2[d\psi +A_a dx^a - \Omega(r)dt]^2 + r^2 \hat{g}_{ab} dx^a dx^b .\nonumber 
\eeqa
Here $f,g,h,\Omega$ are functions of $r$ which can be found in Ref. \cite{Dias:2010eu}, and $x^a, \hat{g}_{ab}$ are the coordinates and Fubini-Study metric on $\mathbb{CP}^N$, respectively. $N$ is related to the spacetime dimension $D$ by $D=2N+3$, while $A_a$ is related to the K\"{a}hler form $J$ by $dA = 2J$. In this equal angular momenta case the above metric functions depend on two dimensionful parameters, which we take to be the horizon radius $r_+$, and the spin parameter $a$. When $a \rightarrow 0$, this reduces to the higher-dimensional Schwarzschild metric, with the $S^{2N+1}$ metric expressed as the Hopf fibration. The presence of the $\mathbb{CP}^N$ factor is what facilitates the separability of the equations into ODE's, rather than PDE's. That this manifold only exists for integer $N$ explains why the separability only happens in odd spacetime dimension. Extremality is reached at $a=a_{\mathrm{ext}}$, the value of which can also be found in Ref. \cite{Dias:2010eu}.
\\
\noindent\indent We studied linear perturbations of this background:
\be g_{\mu\nu} = \bar{g}_{\mu\nu} + h_{\mu\nu}, \ee
where barred quantities refer to the MP background. We then solved the linearized vacuum Einstein equations, which after imposing the traceless and transverse gauge conditions become
\be (\Delta_L h)_{\mu\nu} \equiv  - \bar{\nabla}^\rho \bar{\nabla}_\rho h_{\mu \nu} - 2\bar{R}_{\mu \rho \nu \sigma} h^{\rho  \sigma} = 0. \ee
We further used the stationarity and axisymmetry of the background metric to decompose the time and azimuthal dependence as $h_{\mu\nu} \propto e^{-i(\omega t - m \psi)}$. Here $\omega$ is a complex number which will be  determined numerically, and $m$ is restricted to be an integer. The azimuthal coordinate is $\psi$, and therefore perturbations with $m=0$ are axisymmetric, whilst those with $m \neq 0$ are non-axisymmetric. Lastly, following \cite{Dias:2010eu}, we separated the angular dependence of the perturbation $h_{\mu\nu}$ through the use of charged scalar harmonics on $\mathbb{CP}^N$. As this decomposition is beyond the scope of this article, and results in numerous lengthy equations, we refer the reader to the original article for details \cite{Dias:2010eu}. Each such charged scalar harmonic can be classified by two integers $(\kappa, m$), with $\kappa \ge 0$. We also mention that mode stability for tensor perturbations was shown in \cite{Kunduri:2006qa}, for both axisymmetric and non-axisymmetric modes. The vector case remains to be investigated.
\\
\noindent\indent Our strategy was to look for exponentially growing solutions to the above equations. These are modes with  $\text{Im}(\omega) > 0$. Threshold unstable modes have $\text{Im}(\omega) = 0$, and the spin at which this happens is labelled $a_{\text{crit}}$. The absence of exponentially growing modes does not establish stability, but the  existence of one clearly does imply an instability. We used the physically relevant boundary conditions of ingoing waves at the horizon, and outgoing near infinity. This choice corresponds to studying the quasi-normal mode spectrum of these black holes \footnote{For a review of quasi-normal modes, see Ref. \cite{Berti:2009kk}.}. With the decomposition of perturbations described above, the problem has now been reduced to a quadratic Sturm-Liouville eigenvalue problem for $\omega$ in a coupled system of ODE's. We used a numerical scheme based on spectral methods to solve these equations, see Ref. \cite{Dias:2010eu} for more details \footnote{Actually, Ref. \cite{Dias:2010eu} used an identical numerical scheme, but applied it to a slightly different problem, the study of the Gregory-Laflamme instability for the black string constructed out of these MP black holes.}.
\\
\begin{figure}
\includegraphics[width=\columnwidth]{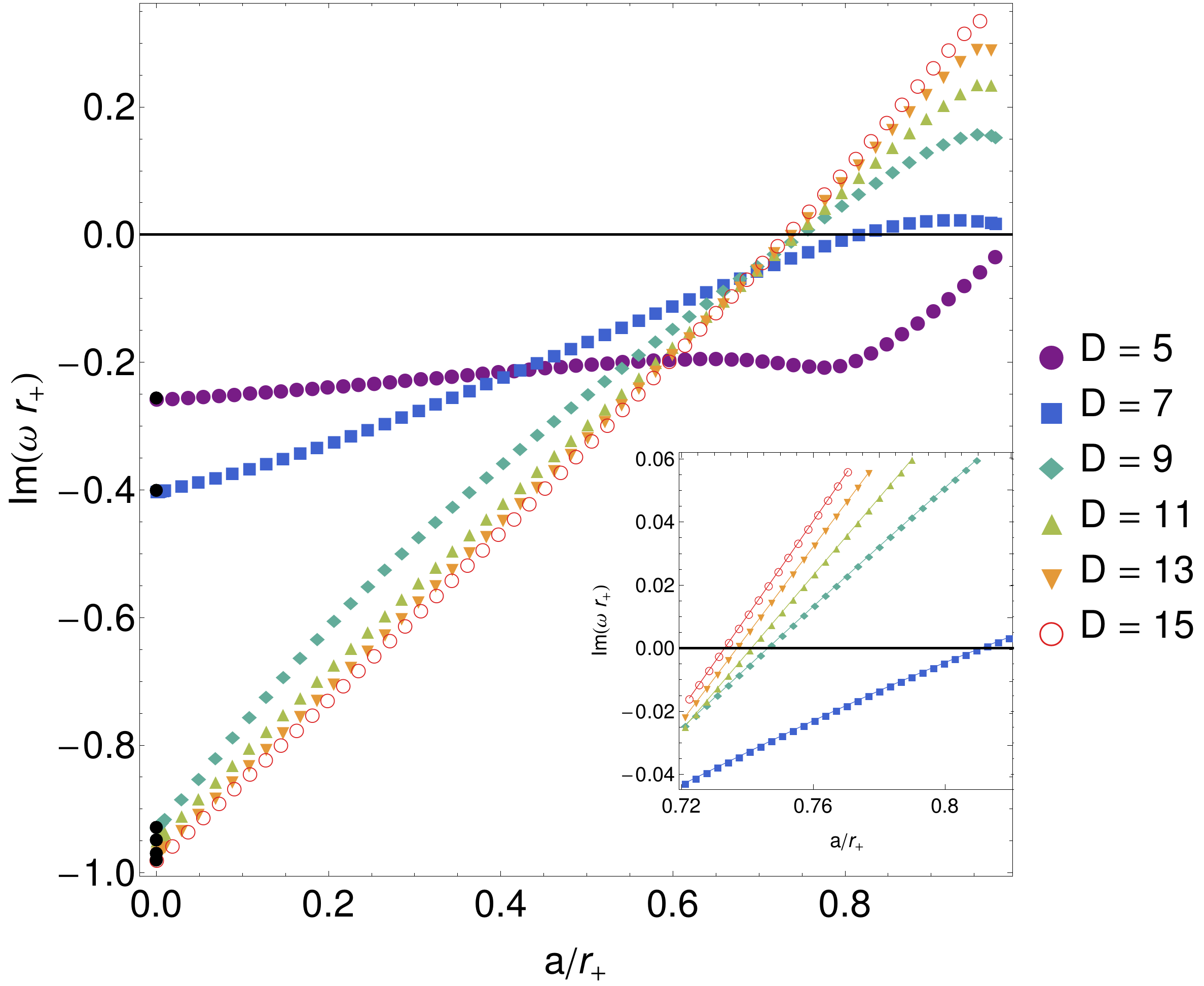}
\caption{Plot of $\text{Im}(\omega)$ for for the dominantly unstable mode, $(\kappa,m) = (0,2)$. The black points at $a=0$ were computed using a different code based upon the gauge invariant formalism of Ref \cite{Kodama:2003jz}. Note that as $D$ increases the critical spin for which $\text{Im}(\omega) = 0$ decreases. For $D=5$, $\text{Im}(\omega) \rightarrow 0^-$ as $a/a_{\text{ext}} \rightarrow 1$, and we find no instability. The inset plot zooms the region where $\text{Im}(\omega)$ becomes positive.}
\label{fig:imaginary}
\end{figure}
\begin{figure}
\includegraphics[width=\columnwidth]{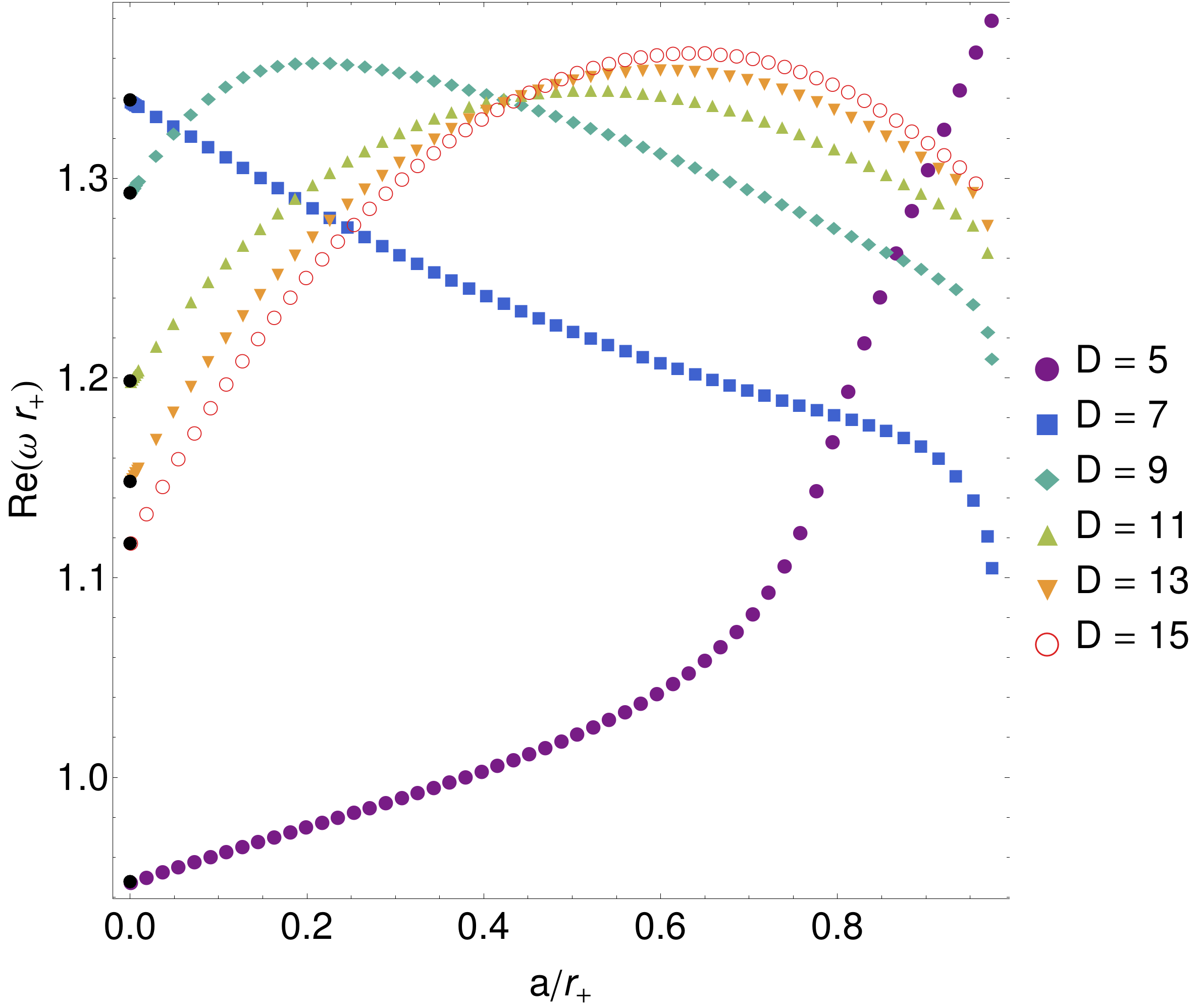}
\caption{Plot of $\text{Re}(\omega)$ for for the dominantly unstable mode, $(\kappa,m) = (0,2)$. The black points at $a=0$ were computed using a different code based upon the gauge invariant formalism of Ref \cite{Kodama:2003jz}.}
\label{fig:real}
\end{figure}
\noindent\indent Our results are as follows. In $D=5$, we find no instability, which is consistent with Ref. \cite{Murata:2008yx}, who also studied the equal angular momenta case. It is also consistent with Ref. \cite{Shibata:2009ad}, who found an instability, but only for singly-spinning black holes. For $D=7,9,11,13,15$, we find numerous bar-mode instabilities. In Fig.~\ref{fig:imaginary} we plot $\text{Im}(\omega)$ for the $(\kappa,m) = (0,2)$ mode, which is the first mode to go unstable as the spin in increased. We refer to this mode as the dominantly unstable mode, as it sets in before any others, and is the mode with the largest growth rate. For completeness we also show in Fig.~\ref{fig:real} the real part of the dominant mode, as a function of $a/r_+$. \noindent\indent 
Our results indicate that $a_{\text{crit}}/r_+$ saturates as a function of $D$, which suggests a possibly analytic understanding in a $1/D$ expansion as in \cite{Emparan:2013moa}.
\\
\noindent\indent We expect that the unstable modes we find will persist for all odd $D \ge 7$. In all dimensions studied, the dominant instability was due to the $(\kappa,m) = (0,2)$ mode. We also found instabilities for $(0,m)$ modes with $m > 2$ as well as for $(1,m)$ modes with $m \ge 1$, but these have a larger value of  $a_{\text{crit}}$  \footnote{This is in contrast to the phenomenon of superradiance in AdS space, where larger $m$ modes become unstable before smaller $m$ modes, first conjectured in \cite{Hawking:1999dp}, and later explicitly checked in \cite{Cardoso:2004hs} using the Teukoslky formalism.} In Table~\ref{tab:jorge} we tabulate the critical spin at which the bar mode and axisymmetric instabilities first set in. The bar mode instabilities set in for much smaller rotation speeds, and are therefore the dominant instabilities of these black holes.
\\
\noindent\indent We computed several checks on our results. First, we computed the Schwarzschild quasi-normal modes using the gauge-invariant formalism of Ref. \cite{Kodama:2003jz} and found that our results reproduced this spectrum as $a \rightarrow 0$. We also computed the axisymmetric $(2,0)$ mode that was first found to be unstable, and compared our result with Ref. \cite{Dias:2010eu}, who only calculated this mode for $\text{Im}(\omega) > 0$. It was expected that as $a$ was increased from zero, $\omega$ would in general be complex, reach $\omega = 0$ at $a_{\text{crit}}$, and then become purely imaginary. However, we find that this mode is always purely imaginary, taking the form $\text{Im}(\omega) = i\,K(a)$ where $K(a)$ is a real function that is negative for $a < a_{\text{crit}}$, zero at $a_{\text{crit}}$, and positive for $a > a_{\text{crit}}$. For $a > a_{\text{crit}}$, our results agreed with Ref. \cite{Dias:2010eu}.
\begin{table}
\centering
\small
\begin{tabular}{|c||c||c|}
\hline
$D$ & $1-a_{\text{crit}}/a_{\text{ext}}$ (NA) & $1 - a_{\text{crit}}/a_{\text{ext}}$ (A) \\
\hline\hline
7 & 0.1891 &  $ 2.339 \times 10^{-5}$ \\
\hline
9 & 0.2537 &  $ 2.116 \times 10^{-3}$ \\
\hline
11 & 0.2587 &  $7.854 \times 10^{-3}$ \\
\hline
13 & 0.2631 & $ 1.504 \times 10^{-2}$ \\
\hline
15 & 0.2669 & $ 2.232 \times 10^{-2}$ \\
\hline
\end{tabular}
\caption{\label{tab:jorge} Critical rotations for non-axisymmetric (NA) and axisymmetric (A) ultraspinning  instabilities. The critical rotation is defined to be the largest rotation such that there are no instabilties for the sector of perturbations in question (axisymmetric or non-axisymmetric). The values for the (A) sector were first presented in \cite{Dias:2010eu} and \cite{Durkee:2010ea}. Note that the bar mode instability sets in for \emph{much} smaller spins.}
\end{table}
\\ \\
\noindent \emph{Discussion:} While a thorough investigation of the linear mode stability is still far from complete for the full MP family, it is nearly finished in the equal angular momenta sector. Scalar perturbations have been examined here in the non-axisymmetric case, and in Ref. \cite{Dias:2010eu} for the axisymmetric one. Tensor perturbations of both types were studied in \cite{Kunduri:2006qa}. Only vector perturbations remain to be analyzed, although it is expected that the dominantly unstable modes will be scalars. Thus it is likely that we have found the dominant instabilities for these black holes, and it is then natural to inquire about the endpoint of this instability.
\\
\noindent\indent Of course, this question deserves a full non-linear numerical treatment, but our results provide some insights. Due to the fact that these perturbations break axisymmetry, the black hole will radiate angular momentum and energy, and in doing so spin down until it reaches a stable spin. In order for the black hole to be able to radiate, the loss of angular momentum and energy must be consistent with Hawking's Area Law, $\delta A \ge 0$. This condition was shown to be equivalent to the superradiant bound $\text{Re}(\omega) - m \Omega_H < 0$ \cite{Shibata:2010wz}, and indeed, we find that only after this condition is satisfied are there modes with $\text{Im}(\omega) > 0$. Therefore, for initial spins slightly larger than the critical values, it is expected that the black holes will simply radiate until they reach a stable configuration. 
\\
\noindent\indent However, for initial spins much larger than the critical value it will take the black holes some finite amount of time to radiate away their excess angular momentum, and during this time the horizon will be rapidly deformed by the growing perturbation. An exciting possible outcome of this scenario would be that the time scale for the growth of the perturbation might exceed the time scale for radiation, and that the black hole might actually fragment into a multiple black hole configuration. If the fragmentation is sufficiently violent, these black holes could fly apart and escape to infinity. Otherwise, they would continue to radiate away energy and angular momentum and would eventually inspiral and merge into a single, non-axisymmetric black hole. This black hole would itself continue to radiate until it settled down to axisymmetric and stationary state. This is a fascinating, but speculative possibility for the endpoint of the instability that would necessarily violate the cosmic censorship.
\\
\noindent\indent This bar mode instability has implications for the stability of some of the more novel black hole solutions that branch off from the MP family. The points in the black hole phase diagram where these solutions join with the MP black holes correspond to  the existence of axisymmetric perturbations with $\omega = 0$. As the spins for which these modes exist are all much greater than the smallest non-axisymmetric $a_{\text{crit}}$ we find, our results suggest that these new solutions will be unstable at least near the branching point, and perhaps more generally.
\\
\noindent\indent A curious feature of our solutions concerned the existence of purely imaginary frequencies $\omega$. As noted earlier, the $(2,0)$ unstable mode was found to be purely imaginary for any $a$. We also found that the $a\rightarrow 0$ limit of the $(1,m)$ unstable modes were purely imaginary. In $D=4$, purely imaginary frequencies have special status, and some of them are associated with changes in the  algebraic classification of the spacetime. The important role that these purely imaginary modes had in determining the stability of the black hole, and in connecting the MP family to new stationary axisymmetric families suggested that there might be a connection with changes in algebraic classifications. However, it has recently been shown that there are no algebraically special modes of Schwarzschild  in $D \ge 5$ \cite{Dias:2013hn}, and therefore it appears unlikely that any special geometrical significance can be assigned to these modes.
\\
\noindent\indent In summary, we have found a new, non-axisymmetric instability of a certain class of higher dimensional rotating black holes. These black holes were previously found to be unstable to axisymmetric perturbations, and the instabilities we find occur for much slower rotation speeds. We expect that the instability we find with the smallest $a_{\text{crit}}$ corresponds to the dominant instability of equal angular momenta Myers-Perry black holes. We discussed two possible endpoints of this instability, either spinning down through gravitational radiation, or through a more complicated process involving black hole fragmentation as an intermediate step.


\vskip 1cm
\centerline{\bf Acknowledgements} 
\noindent It is a pleasure to thank Gary Horowitz, \'Oscar Dias, and Harvey Reall for helpful discussions throughout this project. This work was supported in part by the National Science Foundation under Grant No. PHY12-05500.


\bibliographystyle{JHEP}
\bibliography{MP_EAM_BarMode.bib}

\end{document}